\documentclass[conference]{IEEEtran}
\IEEEoverridecommandlockouts
\usepackage{cite}
\usepackage{amsmath,amssymb,amsfonts}
\usepackage{siunitx}
\usepackage{algorithm,algorithmic}
\usepackage{graphicx}
\usepackage{textcomp}
\usepackage{xcolor}
\usepackage{hyperref}
\usepackage{tabularx, booktabs}
\usepackage{marvosym}
\usepackage{tikz}
\usepackage{cleveref}
\usepackage{pifont, lipsum}
\usepackage{stfloats}
\usepackage[table]{xcolor}
\usepackage{textcase}
\usepackage{tikz}

\def\BibTeX{{\rm B\kern-.05em{\sc i\kern-.025em b}\kern-.08em
    T\kern-.1667em\lower.7ex\hbox{E}\kern-.125emX}}
\begin{document}

\title{Physics-Infused Neural MPC of a DC-DC Boost Converter with Adaptive Transient Recovery and Enhanced Dynamic Stability\\
}

\author{
	\IEEEauthorblockN{Tahmin Mahmud}
	\IEEEauthorblockA{
		Elmore Family School of Electrical and Computer Engineering, Purdue University, West Lafayette, IN 47907, USA \\
		Email: mahmud13@purdue.edu
	}
}

\maketitle

\begin{abstract} 
DC-DC boost converters require advanced control to ensure efficiency and stability under varying loads. Traditional model predictive control (MPC) and data-driven neural network methods face challenges such as high complexity and limited physical constraint enforcement. This paper proposes a hybrid physics-informed neural network (PINN) combined with finite control set MPC (FCS-MPC) for boost converters. The PINN embeds physical laws into neural training, providing accurate state predictions, while FCS-MPC ensures constraint satisfaction and multi-objective optimization. The method features adaptive transient recovery, explicit duty-ratio control, and enhanced dynamic stability. Experimental results on a commercial boost module demonstrate improved transient response, reduced voltage ripple, and robust operation across conduction modes. The proposed framework offers a computationally efficient, physically consistent solution for real-time control in power electronics.
\end{abstract}

\section{Introduction} \label{sec1}
Boost converters are widely adopted in modern power electronic systems (PESs) due to their inherently unidirectional power flow which aligns seamlessly with broad range of applications including PV strings, AI data centers, fuel cells, and EV powertrains \cite{b1}. Evolving beyond traditional DC-link regulation remits, recent control-centric developments prioritize health-aware intelligence. Leveraging data-driven frameworks such as edge AI \cite{b2} and digital twins (DT) \cite{b3}, enable early detection of failure modes and degradation parameters from high-rate control telemetry. This further enhances continuous offline parameter identification, adaptive duty-ratio and controller gain scheduling \cite{b4}, as well as constraint tightening for predictive noise tolerance and reliability-centered maintenance. Advanced data-driven control schemes preserve stability margins, transient performance, and efficiency across wide operating envelopes despite parameter drift and load variability. 

\begin{figure}[htbp]
	\centering
	\includegraphics[width=\columnwidth]{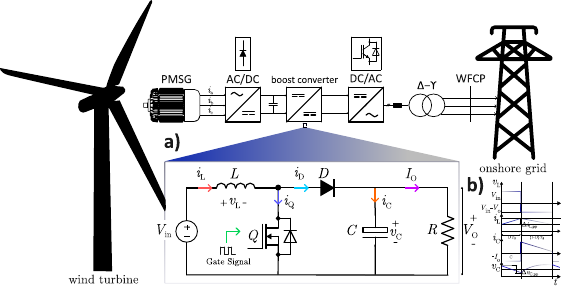} 
	\caption{Power conversion layout for variable-speed WECS. (a) System architecture, and (b) Key waveforms of the boost converter under CCM.}
	\label{fig1}
\end{figure}

In variable-speed wind energy conversion system (WECS) \cite{b5} utilizing permanent magnet synchronous generators (PMSGs) and diode rectifiers as illustrated in Fig.~\ref{fig1}(a), adding a boost stage decouples the generator EMF from the DC-link voltage, thereby allowing active regulation of both generator current and voltage. This widens the variable speed range at low winds, advances torque control for maximum power point tracking (MPPT), and reduces device count and cost compared to full back-to-back (BTB) PWM-assisted topology.

\begin{figure*}[htbp]
	\centering
	\includegraphics[width=\textwidth, keepaspectratio]{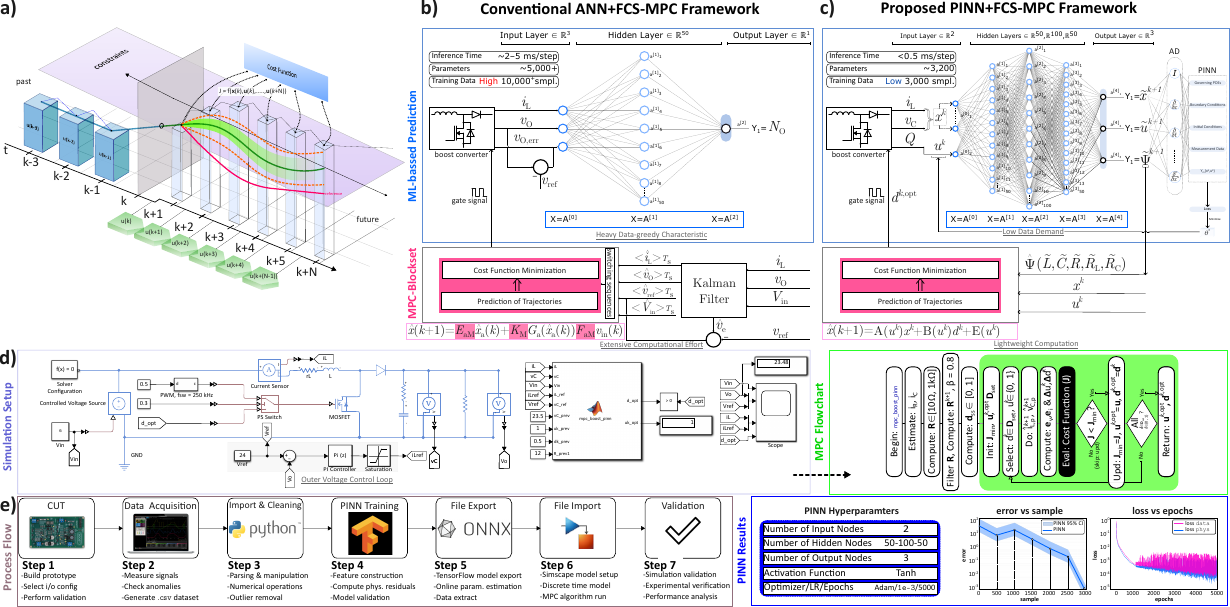}
	\caption{Conceptual illustration of the high-fidelity PINN+FCS-MPC framework. (a) Receding horizon policy, (b) Governing principle of the conventional ANN+FCS-MPC framework, (c) Governing principle of the proposed PINN+FCS-MPC framework, (d) FCS-MPC control structure in MATLAB/Simscape (left-panel) and MPC Flowchart (right-panel), (e) Hierarchical PINN+FCS-MPC offline deployment pipeline (left-panel) and PINN results (right-panel).}
	\label{fig2}
\end{figure*}

Despite its system-level advantages, the boost converter introduces significant nonidealities and nonlinearities that strongly affect control, efficiency, and reliability \cite{b6}. Therefore, accurate modeling and mitigation of non-ideal effects including dead-time ($T_{\mathrm{d}}$), parasitic oscillations, and loss mechanisms are indispensable for a consistent co-existence with emerging learning-enabled control techniques.

Data-driven models, especially Recurrent Neural Networks (RNNs), Multilayer Perceptrons (MLPs), and Deep Neural Networks (DNNs) have emerged as powerful metamodels for PESs, offering fast, parallelizable inference ideally suited for real-time tasks such as online estimation, model predictive control (MPC) \cite{b8}, and edge deployment. They capture complex nonlinear dynamics via universal approximation and enable fast inference, but purely data-driven approaches suffer from poor out-of-distribution generalization, brittleness, high data demands, and error accumulation in closed-loop safety-critical power electronics applications \cite{b9}.

To overcome these shortcomings, contemporary research practices have been shifted towards hybrid data-driven modulation schemes. Physics-informed neural networks (PINNs) are an emerging paradigm for forward and inverse PDE-constrained problems in system-level PES. They are directly embedded with physical laws (ordinary differential equations (ODEs)/ partial differential equations (PDEs), circuit equations, boundary/initial conditions) as constraints in the loss function \cite{b10}. This physics-based regularization drastically reduces reliance on labeled data, constrains the solution space to physically feasible operating regimes, and significantly improves extrapolation, robustness, and data efficiency \cite{b11}.

The literature on advanced boost converter control can be categorized into AI/ML-based optimization methods, predictive and disturbance-rejection schemes for constant power loads (CPLs), hybrid nonlinear control techniques, and digitally implemented high-performance controllers for FPGAs or MCUs. This study focuses exclusively on data-driven approaches. As reported in \cite{b13}, the authors propose a reinforcement learning (RL)-based nonlinear controller for a DC–DC boost converter that replaces a heavy DNN policy with a lightweight regression-based policy function, enabling on-device MCU implementation with low compute. However, the learned regression policy is a data-fit surrogate of duty-ratio versus load impedance and does not embed the converter’s governing ODEs. In \cite{b14}, a physics‑informed BiLSTM (BiLSTM‑PINN) surrogate for time‑domain simulation of a closed‑loop boost converter is introduced. The proposed framework integrates a power‑balance physics loss, and it is rigorously benchmarked against fully-connected NN (FCNN) and BiLSTM baselines. However, the proposed simulator targets averaged‑model prediction rather than control synthesis. In \cite{b15}, a composite nonlinear controller for a boost converter feeding CPLs based on nonlinear disturbance observer (NDO) and backstepping algorithm has been proposed. But, the method lacks a horizon‑based optimization framework, and it relies on model accuracy and high‑gain tuning. A composite voltage-regulation strategy is proposed in \cite{b16} that fuses PINN-based online estimation of disturbances with a Lyapunov-based backstepping controller to stabilize a boost converter feeding CPLs. However, this scheme depends on averaged dynamics and offers limited mechanisms to enforce input/state constraints under rapid operating changes. While PINN-based modeling has been widely explored for various power electronics topologies, such as grid-following converters (GFLs) \cite{b17}, dual-active-bridge (DAB) \cite{b18}, 3P voltage-source inverters (VSIs) \cite{b19}, and resonant converters \cite{b20}. None of these studies have incorporated a finite receding-horizon based optimization technique.

A PINN model excels at modeling the nonlinear dynamics of a boost converter and respect physical constraints during training, but it operates as an implicit model to systematically enforce hard operational constraints at runtime. Integrating a receding-horizon based policy, such as Finite Control Set MPC (FCS-MPC), with the PINN as the predictor enables systematic constraint satisfaction and explicit multi-objective trade-off management, making the hybrid approach essential for safe and high-performance boost converter control. Unlike, conventional data-driven FCS-MPC \cite{b21} or standalone PINN, this work proposes a simpler yet effective hybrid framework, built upon the methodology presented in \cite{b22}. The critical contributions of this paper are summarized below:
\begin{enumerate}
	\item Development of a lightweight PINN architecture with embedded physical constraints for state prediction. Adaptable to DAB, 3P VSI \& other DC-DC converters.
	\item Integration of PINN+FCS-MPC explicit multi-objective optimization and robust load transient compensation.
	\item Experimental validation of improved transient recovery and dynamic stability on a commercial boost converter platform.
\end{enumerate}

Subsequent sections are organized as follows: Section \ref{sec2} reviews the foundational concepts, Section \ref{sec3} presents the proposed hybrid PINN+FCS-MPC framework and its modeling, Section \ref{sec4} validates the approach through a case study and experimental results, Section \ref{sec5} concludes the paper.

\section{Preliminaries} \label{sec2}

\subsection{Finite Control Set Model Predictive Control (FCS-MPC)}
MPC is an advanced optimal control strategy that explicitly utilizes a dynamic model of the system to predict the future behavior over a receding horizon and computes the control action by solving a constrained optimization problem at each sampling instant as illustrated in Fig.~\ref{fig2}(a). Among the MPC variants applied to power electronics and drives, FCS-MPC outperforms Continuous Control Set MPC (CCS-MPC) by directly exploiting the discrete nature of converters, eliminating the need for modulators. FCS-MPC enforces constraints via simple enumeration, offering simpler implementation, higher robustness, and better performance under uncertainties and nonlinearities \cite{b23}.

\subsubsection{Prediction Model}
The basic prediction model used in the FCS-MPC implementation for the boost converter is the discrete-time (discT) state-space model obtained by forward Euler discretization (\ref{eq1}) of the continuous-time plant dynamics.

\begin{equation} \label{eq1}
	\dot x(t_k)\Delta\mathrm{t} = x^{k+1}-x^{k}
\end{equation}
\vspace{-0.6cm}
\begin{equation} \label{eq2}
	\dot x = \textbf{A}x+\textbf{B}u+f(\zeta), \ \text{and} \ y = \textbf{C}x+ \textbf{D}u
\end{equation}

Eq. (\ref{eq2}) is the continuous-time linearized state-space model. Where, \(x\) represents the PES physical input state variable, \(u\) as control state variable and \(y\) as output state variable. \(f(\zeta)\) is the piecewise-constant affine term (duty-ratio, \(d\)) in the continuous-time dynamics that acts on the system independently of the PES state and controlled input variables. \textbf{A}, \textbf{C} and \textbf{B}, \textbf{D} are the continuous-time physical input state and control state matrices respectively. Now, \(x = \begin{bmatrix} \vec{i_{\mathrm{L}}} & \vec{v_{\mathrm{C}}} \end{bmatrix}^{\!\top}
\), having multiple state variables shifts the discretization problem from first-order to a second-order state-space discretization challenge. Therefore, over one sampling period $[k_\mathrm{i}T_{\mathrm{s}}, k_\mathrm{i+1}T_{\mathrm{s}})$:, instead of using the forward-Euler  discretization, we apply \textit{Heun's} method which yields,
\vspace{-0.1cm}
\begin{equation} \label{eq3}
	\left\{
	\begin{aligned}
		\hat{x}^{k+1} &= \underbrace{(I + T_{\mathrm{s}} \mathbf{A})}_{\mathbf{A}_{\mathrm{j}}}
		 x^k 
		+ \underbrace{T_{\mathrm{s}} \mathbf{B}}_{\mathbf{B}_{\mathrm{j}}} u^k 
		+ \underbrace{T_{\mathrm{s}} \mathbf{f}}_{\mathbf{C}_{\mathrm{j}}} \zeta^k \\[4pt]
		x^{k+1}      &= \hat{x}^{k+1} + \frac{T_{\mathrm{s}}}{2} \mathbf{A} (\hat{x}^{k+1} - x^k)
	\end{aligned}
	\right.;\\\
	\begin{aligned}
		&x^k \in [\underline{x},\, \bar{x}] \\
		&u^k \in [\underline{u},\, \bar{u}] \\
		&\zeta^k \in \mathbf{Z} \\
		&\forall k = 0,1,2\dots
	\end{aligned}
\end{equation}
\vspace{-0.4cm}

\noindent where, $T_{\mathrm{s}}$ represents the sampling interval for discretization. $x^k$, $u^k$ and $\zeta^k$ denote the PES state variables, inputs and model parameters at the $k^{th}$ sampling instant. During each sampling interval, the model parameters $\zeta$ along with the initial conditions are updated. \(I\) is the identity matrix, ${\mathbf{A}_\mathrm{j}}$, ${\mathbf{B}_\mathrm{j}}$ and ${\mathbf{C}_\mathrm{j}}$ are the discrete-time PES state, input and model parameter matrices. $\hat{{x}}^{k+1}$ represents the predictor-corrector of state variables, and $x^{k+1}$ signifies the predicted state variables at the $(k+1)^{th}$ sampling instant. The final prediction model for a boost converter can be expressed by (\ref{eq4}) which will be discussed more in detail in Section \ref{sec3}.
\vspace{-0.05cm}
\begin{equation} \label{eq4}
	\begin{aligned}
		x^{k+1} &= \mathbf{\Phi}x^k + \mathbf{\Gamma}u^k + \mathbf{\Xi}\zeta,\quad 
		y^k = \mathbf{\Upsilon}x^k + \mathbf{\Pi}u^k;\\
		&\text{where, }\mathbf{\Phi}_{n\times m},\;\mathbf{\Gamma}_{n\times n},\;\mathbf{\Xi}_{n\times o},\;\mathbf{\Upsilon}_{n\times p},\; \mathbf{\Pi}_{n\times q}\;
	\end{aligned}
\end{equation}

\subsubsection{Cost Function Minimization}
The cost function's primary goal is to evaluate all feasible switching states \(u^k\) of the boost converter and select the one that yields the smallest prediction error. Following standard FCS-MPC literature, this cost is interchangeably denoted as \(g\) or \(J\). At each sampling instant, the controller computes this cost and directly applies the optimal switching state \(u^{k,\mathrm{opt}}\), that minimizes it, thereby ensuring accurate reference tracking. 
\begin{equation} \label{eq5}
g = (y^{k+1}_{\mathrm{ref}} - \hat{y}_{\mathrm{p}}^{\,k+1})^2\\
\end{equation}

In real-time operation, \(u^k\) and $g$ are determined at step $k$, the controller predicts the system state at $k+1$ and computes the optimal input as,
\vspace{-0.1cm}
\begin{equation} \label{eq6}
 u^{k,\mathrm{opt}} = \underset{u^k \in \mathcal{U}}{\operatorname{arg\,min}} \; g^{k+1}(u^k)\\
\end{equation}
  with $\mathcal{U}$ denoting the admissible input set.

\subsubsection{Multi-Objective Terms \& Weighting Factor Design}
In MPC, the cost function $J$ is minimized to track the reference \(y^{k+1}_{\mathrm{ref}}\) while considering additional objectives. Multiple terms can be introduced in $J$, e.g., capacitor voltage/inductor current reference tracking, switching frequency reduction etc. with weighting factor $\lambda$ balancing their influence: 
\begin{equation} \label{eq7}
	\begin{aligned}
		J = 
		\underbrace{(y^{k+1}_{\mathrm{ref}} - \hat{y}_{\mathrm{p}}^{\,k+1})^2}_{\text{primary term}} 
		+ 
		\underbrace{\lambda\, \Delta z^2}_{\text{secondary term with weighting factor } \lambda} \\[2mm]
		\hspace{1cm}\text{where,} \ \Delta z = z^k - z^{k-1},\ \text{and}\ \lambda \in [0, 10]
	\end{aligned}
\end{equation}

\subsection{Physics-Informed Neural Network (PINN)}
\subsubsection{The Building Blocks of PINNs}
The PINNs adopt a fully connected feed-forward NN (FFNN) architecture also known as a standard MLP. At each control instant, the proposed network receives the spatiotemporal information through an input layer, \(\mathbf{a}^{(0)} = [x^k, u^k]\) of dimension equal to the state-plus-control action space. This information is then propagated through a sequence of densely connected hidden layers with nonlinear activations \(\sigma\) (typically \texttt{tanh} or \texttt{ReLU}), yielding a high-dimensional latent representation. Successive layers $\ell$ are equipped with learnable weights 
$\mathbf{W}^{(\ell)} \in \mathbb{R}^{d_{\ell-1} \times d_{\ell}}$ 
and biases $\mathbf{b}^{(\ell)} \in \mathbb{R}^{d_{\ell}}$, 
where $d_{\ell}$ denotes the output size of the current layer $\ell$. 
The hidden layers \(\mathbf{a}^{(\ell)}\) and the final output layer \(\mathbf{a}^{(L)}\) follow the standard recursive computation scheme \cite{b24} detailed below in Eq. (\ref{eq8}).
\vspace{-0.03cm}
\begin{equation} \label{eq8}
	\begin{aligned}
		\mathbf{a}^{(\ell)} &= \sigma \thinspace \!\bigl( \mathbf{W}^{(\ell)} \mathbf{a}^{(\ell-1)} + \mathbf{b}^{(\ell)} \bigr), 
		&& \hspace{-0.2cm} \ell = 1,\dots,L-1 \ ; \sigma = \texttt{tanh}\\[6pt]
		\mathbf{a}^{(L)}     &= \mathbf{W}^{(L)} \mathbf{a}^{(L-1)} + \mathbf{b}^{(L)}, 
		&& \hspace{-0.2cm} \ell = L
	\end{aligned}
\end{equation}
\vspace{-0.2cm}
\noindent such that, the FFNN can be defined as,
\vspace{-0.05cm}
\begin{equation} \label{eq9}
\tilde{\boldsymbol{\kappa}}_{\boldsymbol{\theta}} = (\mathcal{A}^{(L)}\circ\mathcal{A}^{(L-1)}\dots\mathcal{A}^{(0)})
\end{equation}

\noindent with function composition, \(\circ\) and trainable parameters, 
$\boldsymbol{\theta} = \bigl\{ \mathbf{W}^{(\ell)}, \mathbf{b}^{(\ell)} \bigr\}_{\ell=1}^{L}$. The final layer produces the approximate solution  to the underlying partial differential equation that governs the converter dynamics, which can be denoted as,

\begin{equation}\label{eq10}
	\underbrace{\tilde{\boldsymbol{\kappa}}_{\boldsymbol{\theta}}(\hat{x}^{k}, \hat{u}^{k}, \hat{\boldsymbol{\Psi}}^{k})}_{\in \mathbb{R}^{3}}
	=
	\begin{bmatrix}
		\tilde{x}^{k+1} \\[4pt]
		\tilde{u}^{k+1} \\[4pt]
		\tilde{\Psi}^{k+1}
	\end{bmatrix}
	=
	\mathcal{A}_{\boldsymbol{\theta}}\bigl(x^{k}, u^{k} \bigr)
\end{equation}

\subsubsection{Ordinary Differential Equations (ODEs)}
PINNs embed the PES’s governing ODEs directly into the training process, turning a generic MLP network into one that learns physically consistent dynamics rather than a black-box model. The system evolution is written as an NN-parameterized ODE:
\begin{equation}\label{eq11}
	\dot{\mathbf{m}}(t) 
	= \boldsymbol{f}\bigl( \mathbf{m}(t), t; \boldsymbol{\theta} \bigr),
	\qquad \mathbf{m}(t_0) = \mathbf{m}_0
\end{equation}
with automatic differentiation (AD) that provides the necessary partial derivatives with respect to both the \(x^k\) and \(u^k\), enabling direct inclusion of continuous-time physics.

\subsubsection{Partial Differential Equations (PDEs)} ODE-PINNs constrain dynamics only in 1-D trajectories, whereas PDE-PINNs enforce physics in space and time, enabling spatially dependent systems. In \cite{b25}, the steady-state PDE is defined as,
\vspace{-0.2cm}
\begin{equation} \label{eq12}
	\begin{pmatrix}
		\tilde{\boldsymbol{\kappa}}_{\boldsymbol{\theta}}(x_{\theta};\mu) - f_y \\[4pt]
		\mathcal{B}\!\left(x_{\theta}\right)
	\end{pmatrix}
	=
	\begin{pmatrix}
		0 \\ 0
	\end{pmatrix},
	\qquad 
	\theta \in \Omega \cup \partial\Omega .
\end{equation}

\noindent
where $\Omega \subset \mathbb{R}^d$ is the spatial domain of dimension $d$, 
$\partial\Omega$ is its boundary, 
$\boldsymbol{\kappa}_{\boldsymbol{\theta}}$ is a (integro-)differential operator with parameters $\mu$, 
and $f_y$ is a forcing term.

\subsubsection{Loss Function}
The loss function is composed of two complementary terms, (a) a data-fitting term that drives the network to fit any available measurements and (b) a physics-informed term that enforces the governing equations of the system. It is defined as,
\begin{equation} \label{eq13}
	\mathcal{L}
	= \mathcal{L}_{\texttt{data}}
	+ \mathcal{L}_{\texttt{phys}}
\end{equation}

\begin{figure*}[!b]
	\vspace{-1\baselineskip}
	\noindent\makebox[\textwidth][l]{\rule{\textwidth}{0.4pt}} 
	\vspace{-2\baselineskip}

	\begin{equation}\label{eq14}
		\footnote{1}
		{\footnotesize
			\begingroup
			\thinmuskip=0.2mu \medmuskip=0.3mu \thickmuskip=0.3mu 
			\relax
			\mathcal{L} = \underbrace{
				w_d\,\frac{1}{|\mathcal{T}_d|}\sum_{\kappa\in\mathcal{T}_d}\bigl|\tilde \kappa_{\theta}(\hat{x}^{k}, \hat{u}^{k}, \hat{\boldsymbol{\Psi}})-\kappa({x}^{k}, {u}^{k}, {\boldsymbol{\Psi}})\bigr|^2}_{\mathcal{L}_{\texttt{data}}}
			+ \underbrace{
				w_f\,\frac{1}{|\mathcal{T}_f|}\sum_{x_f\in\mathcal{T}_f}\bigl|\mathcal{F}(x_f,t;\theta)\bigr|^2
				+
				w_b\,\frac{1}{|\mathcal{T}_b|}\sum_{x_b\in\mathcal{T}_b}\bigl|\tilde u(x_b,t;\theta)-g(x_b,t)\bigr|^2
				+
				w_i\,\frac{1}{|\mathcal{T}_i|}\sum_{x_i\in\mathcal{T}_i}\bigl|\tilde u(x_i,0;\theta)-h(x)\bigr|^2}_{\mathcal{L}_{\texttt{phys}}}
			\endgroup
		}
	\end{equation}
	\vspace{-1\baselineskip}
\end{figure*}

In Eq. (\ref{eq14}), the PINN loss function is expressed in its expanded form. Here, $w_f$, $w_b$, and $w_i$ denote the weights, while $T_f$, $T_b$, and $T_i$ represent the sets of samples from the interior domain, boundary, and initial conditions, respectively.

\section{Methodologies: Hybrid PINN+FCS-MPC} \label{sec3}
In contrast to the conventional data-intensive and complex ANN+FCS-MPC architecture as shown in Fig.~\ref{fig2}(b), the proposed PINN+FCS-MPC framework as illustrated in Fig.~\ref{fig2}(c) adopts a lightweight and physics-dominant architecture tailored to the boost converter. The hybrid scheme comprises three tightly integrated modules: (a) the (minimal) data-driven step, (b) the physics-informed residual enforcement step, and (c) the FCS-MPC layer, whose formulation and interaction are detailed in the following subsections.

\subsubsection{\textnormal{\texttt{data}}-\textnormal{driven} Step}
The \({\texttt{data}}\)-driven step is a compact recurrent neural surrogate. 
At each instant, the module receives $\{x^{k-i}\}_{i=0}^{3}$ and $\{u^{k-i}\}_{i=0}^{3}$ 
(with switching state $u^k=\mathbf{1}_{\{Q=1\}}$ for the controller blockset, and $d\in[0,1]$ for averaged models).

To enhance robustness over the full operating envelope, different operational cycle parameters are embedded within the MLP network as trainable constants. In addition to this, we normalize the sampling period $T_{\mathrm{s}}$ and keep it phase-aligned within the switching cycle. The network outputs the one-step-ahead state prediction matrices,
\vspace{-0.5cm}
\begin{equation} \label{eq15}
	\tilde{x}^{k+1} =
	\begin{bmatrix}
		\tilde{i}_\mathrm{L}^{k+1}\\[4pt]
		\tilde{v}_\mathrm{C}^{k+1}
	\end{bmatrix},
	\
	\tilde{u}^{k+1} =
		\tilde{Q}_{\mathrm{ON/OFF}}^{k+1}\\[5pt],
	\
	\tilde{\boldsymbol{\Psi}}^{k+1}\!=\!
	\begin{bmatrix}
		\tilde{L}\\[1pt]
		\tilde{C}\\[1pt]
		\tilde{R}\\[1pt]
		\tilde{R}_\mathrm{L}\\[1pt]
		\tilde{R}_\mathrm{C}
	\end{bmatrix}
\end{equation}

At runtime, ultra-low-latency predictions of \(\hat{x}^k\) and \(\hat{u}^k\) are directly fed to the \texttt{phys} module and \(\hat{\boldsymbol{\Psi}}^{k}\) to the FCS-MPC layer. Thus, we adopt a one-step prediction horizon and exploit online estimates of the physical parameters to continuously update the discrete-time system $\boldsymbol{\Phi}$, $\boldsymbol{\Gamma}$, $\boldsymbol{\Xi}$ and output matrices $\boldsymbol{\Upsilon}$, $\boldsymbol{\Pi}$ respectively. The $\mathcal{L}_{\texttt{data}}$ metric measures the deviation between the PINN prediction $\tilde{\cdot}^{k+1}$ and the true state ${\cdot}^{k+1}$ at the next time step $k+1$. The \texttt{data} loss is calculated as, \\
\vspace{-0.2cm}
{\small \(\textsf{RMSE,}\,\mathcal{L}_{\texttt{data}} \\[2pt]\)}
{\small
	\begin{equation}\label{eq16}
		\begin{alignedat}{2}
			= \sqrt{\frac{1}{N}
				\sum_{i=1}^{N}\Bigl[
				(\Delta i_{\mathrm{L}}^{k+1})^{2}
				+ (\Delta v_{\mathrm{C}}^{k+1})^{2}
				+ (\Delta Q^{k+1})^{2}
				+ (\Delta \Psi^{k+1})^{2}
				\Bigr]}
			\\[-2pt]
			\text{\small where } \Delta(\cdot)=\tilde{(\cdot)}-(\cdot)
		\end{alignedat}
	\end{equation}
}

\subsubsection{\textnormal{\texttt{phys}}-\textnormal{driven} Step}

\texttt{phys} module functions as a physics-based temporal regularizer that embeds the converter's discrete-time governing equations as a soft constraint within the NN's objective function. It ensures that the network's predicted state evolution over a time step ($\Delta t$) aligns with the boost-integrated PES's governing equations. The continuous-time dynamics $\dot{m} = \mathbf{f}(m, u)$ are approximated as,
\vspace{-0.1cm}
\begin{equation} \label{eq17}
	\begin{aligned}
		\textsf{fwd (phys residual): } & m^{k+1}\!=\!m^k + \Delta t\, \cdot \mathbf{f}(m^k,u^k),\\
		\textsf{bwd (synthDA): } & m^k\!=\!m^{k+1} - \Delta t\, \cdot \mathbf{f}(m^{k+1},u^{k+1})
	\end{aligned}
\end{equation}
\vspace{-0.4cm}
\\
where, \textsf{fwd}-step enforces ODE consistency at inference and \textsf{bwd}-step produces 2$\times$ synthetic training samples by reverse-integrating from measured states w/o any new iterations. Here,
\[
\mathbf{f}(m^k, u^k) =
\begin{pmatrix}
	\dfrac{1}{L} \Bigl( V_\mathrm{in}' - (1-\bar{d})\,v_\mathrm{C}' + \bar{v}_\mathrm{C}\,d' \Bigr) \\[4pt]
	\dfrac{1}{C} \Bigl( (1-\bar{d})\,i_\mathrm{L}' - \dfrac{v_\mathrm{C}'}{R} - \bar{i}_\mathrm{L}\,d' \Bigr)
\end{pmatrix}
\
\text{\tiny 
	\begin{tabular}{@{}l@{}}
		$'$ = small-signal pert. \\ 
		$\bar{\cdot}$ = steady-state
\end{tabular}}
\]

The \texttt{phys} stage computes the physics residual based on \textsf{fwd}-step via substituting the network-predicted states
\(m^{\,k+1}_{\theta}\) into the discrete update equation. Eq. (\ref{eq18}) enforces the forward direction and penalizes \(m^{\,k+1}_{\theta}\), under any violation of the discretized ODE given \(m_{\theta}^{\,k}\ \text{and} \ u^k\).
The resulting \texttt{phys} loss:
\vspace{-0.1cm}
{\small
	\begin{equation}\label{eq18}
		\begin{aligned}
			\textsf{RMSE,}\;\mathcal{L}_{\texttt{phys}}
			&=
			\tfrac{1}{N} \sum_{k=1}^{N} 
			\bigl|\bigl| m^{\,k+1}_{\theta} - m^{\,k}_{\theta} - \Delta t\,\cdot \mathbf{f}(m_{\theta}^{\,k}, u^k) \bigr|\bigr|^2
		\end{aligned}
	\end{equation}
}

\subsubsection{FCS-MPC Layer}
The FCS-MPC layer is programmed and deployed in MATLAB/Simscape environment as illustrated in the  Fig.~\ref{fig2}(d), left-plane. The FCS-MPC layer employs an outer voltage control loop to generate \(i_\mathrm{L,ref}^{k+1}\) and an inner current control loop with discrete-time predictor updated via PINN-informed parameters \(\hat{\boldsymbol{\Psi}}^{k}\). 
The discrete-time model of the boost converter is described in Eq. (\ref{eq4}),

with, 
\(x^k = \begin{bmatrix} i_\mathrm{L} \\ v_\mathrm{C} \end{bmatrix},\;
y^k = \begin{bmatrix} i_\mathrm{L} \\ V_\mathrm{O} \end{bmatrix},\;
u^k = \begin{cases} 1 & Q=1\\ 0 & Q=0 \end{cases}\)

The system and output matrices can be expressed as,
\begin{equation} \label{eq19}
\begin{aligned}
	&\smash{\text{Let, }\ \mathcal P\!=\!T_{\mathrm s}R_{\mathrm t}^{\scriptscriptstyle -1}, \mathbb Q\!=\!RL^{\scriptscriptstyle -1},\mathfrak R\!=\!C^{\scriptscriptstyle -1}, \mathsf S\!=\!R_{\mathrm C}R_{\mathrm t}^{\scriptscriptstyle -1}, \mathbb T\!=\!RC^{\scriptscriptstyle -1}}\\
	&\mathbf{\Phi}\!=\![1-\tau T_{\mathrm s} (R_\mathrm{L}L^{\scriptscriptstyle -1}+\mathbb Q\mathsf S) \ -\tau\mathcal P\mathbb Q, \ \ \tau\mathbb T\mathcal P \ \ 1-\mathfrak R\mathcal P]^{\mathsf T},\\
	&\mathbf{\Gamma}\!=\!\mathcal P[\mathbb Q(R_{\mathrm C}i_{\mathrm L}+ v_{\mathrm C}),-\mathbb T i_{\mathrm L}]^{\mathsf T},\quad
	\mathbf{\Xi}\!=\!(\tau T_{\mathrm s}L^{-1})[1,0]^{\mathsf T},\\
	&\mathbf{\Upsilon}\!=\!\operatorname{diag}(1,0)+R\,[0,1]^{\mathsf T}[\tau\mathsf S\ \ R_{\mathrm t}^{-1}]^{\mathsf T},\ \ 
	\mathbf{\Pi}\!=\![0,-\mathsf S R\,i_{\mathrm L}]^{\mathsf T}\\
	& \text{where, }R_{\mathrm t}\!=\!R+R_{\mathrm C}
\end{aligned}
\end{equation}

In Eq. (\ref{eq19}), the symbol \(\tau\) is an auxiliary scalar (mode indicator) that distinguishes how the boost converter's inductor current, \(i_\mathrm{L}\) evolves within one sampling period \(T_\mathrm{s}\) to capture the CCM/DCM transition. In addition to this, the four operating modes of the boost converter defined by the control signal \(u^k\), \cite{b26} are: (a) $u^k=1$, $\tau=1 \Rightarrow$ $Q$ ON, CCM; (b) $u^k=0$, $\tau=1 \Rightarrow$ $Q$ OFF, CCM; (c) $u^k=0$, $\tau=\tau^*/T_{\mathrm s} \Rightarrow$ $Q$ OFF, transient between CCM and DCM with $\tau^*$ from $\hat{i}_\mathrm{L,p}^{k+1}=0$; (d) $u^k=0$, $\tau=0 \Rightarrow$ $Q$ OFF, DCM. Furthermore, the one-step ahead predicted inductor current, \(i_\mathrm{L}\) can be derived as,
\vspace{-0.2cm}
\begin{equation}  \label{eq20}
	\hat{i}_\mathrm{L,p}^{k+1} = \!(1-\tau^*)i_\mathrm{L}^k [R_\mathrm{L}L^{\scriptscriptstyle -1}+\mathbb Q\mathsf S]-v^k_\mathrm{C}(\mathbb Q \times R_\mathrm{t}^{-1}) + \tau^*\frac{V^k_\mathrm{in}}{L}
\end{equation}

In Eq. (\ref{eq20}), when $\hat{i}_\mathrm{L}^{k+1}=0$, the $\tau^*$ can be obtained by,

\begin{equation} \label{eq21}
\tau^{*}
= \frac{
	i_\mathrm{L}^{k+1}
	- i_\mathrm{L}^{k}\!\left(R_{\mathrm{L}}L^{-1}+\mathbb{Q}\mathsf S\right)
	+ v_{\mathrm{C}}^{k}\!\left(\mathbb{Q} R_{\mathrm{t}}^{-1}\right)
}{
	\dfrac{V_{\mathrm{in}}^{k}}{L}
	- i_\mathrm{L}^{k}\!\left(R_{\mathrm{L}}L^{-1}+\mathbb{Q}\mathsf S\right)
}
\end{equation}

\begin{figure*}[ht]
	\centering
	\includegraphics[width=\textwidth]{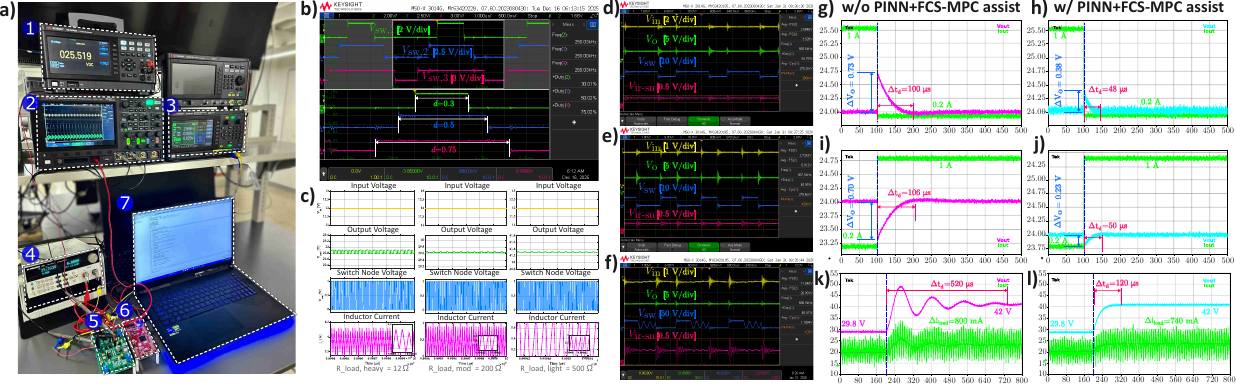}
	\caption{Experimental setup and performance evaluation of the CUT with PINN+FCS-MPC assisted control. 
(a) Prototype testbed (\(\textcircled{1}\) DMM, \(\textcircled{2}\) O-scope, \(\textcircled{3}\) PSU, \(\textcircled{4}\) DC EL, \(\textcircled{5}\) TI-PMLKBOOSTEVM, \(\textcircled{6}\) TI-$\mu$C, and \(\textcircled{7}\) host PC), (b) Measured $V_{\mathrm{sw}}$ under different duty ratios, (c) Simulated steady-state operation under CCM$\dashrightarrow$BCM$\dashrightarrow$DCM, (d)–(f) Experimental steady-state waveforms under constant-current loads of 0.6~A, 0.4~A, and 0.3~A, (g)–(j) Output voltage TR to load current steps from 1~A to 0.2~A and from 0.2~A to 1~A, w/o and w/ PINN+FCS-MPC assist, (k)–(l) Load TR in CR mode: $200 \ \Omega \dashrightarrow 500 \ \Omega$.}
	\label{fig3}
\end{figure*}

\vspace{-0.1cm}
At each sampling instant $k$, the proposed FCS-MPC framework jointly optimizes two control variables: the binary switching state $u^k \in \{0,1\}$ representing the ON/OFF status of the bottom switch, and a continuous duty ratio $d^k \in [0,1]$ that commands the pulse-width modulation (PWM) stage. The controller operates on the measured state vector $x^k$. For computational tractability within the FCS-MPC paradigm, the continuous duty ratio is discretized into a finite admissible set $\mathcal{D} = \{d^{(1)}, d^{(2)}, \ldots, d^{(M)}\} \subset [0,1]$;  $\lvert\mathcal{D}\rvert = M \Rightarrow 2M = 20$ cost function iterations per $T_{\mathrm{s}}= \ $1.67 $\mu$s. The one-step-ahead state prediction is obtained via the PINN-augmented discT:
\begin{equation} \label{eq22}
	\tilde{x}^{k+1}
	=
	f_\mathrm{d}\!\left(
	x^{k},
	u^{k},
	d^{k},
	V_{\mathrm{in}}^{k},
	R^{k+1}
	\right)
\end{equation}
yielding predicted inductor current $\hat{i}_\mathrm{L,p}^{k+1}$ and capacitor voltage $\hat{v}_\mathrm{C,p}^{k+1}$. The model explicitly incorporates the anticipated load resistance $R^{k+1}$, enabling proactive compensation for load transients. The capacitor current is approximated in discrete-time using a backward-Euler scheme. Applying the current balance at the output node,
$i_{\mathrm{L}} = i_{\mathrm{C}} + i_{\mathrm{R}}$, the load current can be expressed as,
\vspace{-0.3cm}
\begin{equation} \label{eq23}
	i_{\mathrm{R}}^{k}
	=
	i_{\mathrm{L}}^{k}
	-
	C \frac{v_{\mathrm{C}}^{k} - v_{\mathrm{C}}^{k-1}}{T_{\mathrm{s}}}
\end{equation}

For a resistive load, the load current satisfies \(i_{\mathrm{R}}^{k} = {v_{\mathrm{C}}^{k}}/{R}\) from which an instantaneous estimate of the load resistance can be obtained at the current sample \(k\),
\vspace{-0.1cm}
\begin{equation} \label{eq24}
	{R}^k
	= \frac{v_\mathrm{C}^{k}}{i_\mathrm{L}^{k} - C \frac{v_\mathrm{C}^{k} - v_\mathrm{C}^{k-1}}{T_\mathrm{s}}}
\end{equation}

To improve robustness against measurement noise and modeling inaccuracies, a first-order recursive filter is applied to ${R}_k$. The resistance estimate used in the predictive controller is then updated according to,
\vspace{-0.1cm}
\begin{equation} \label{eq25}
	R^{k+1} = \beta\,R^k + (1-\beta)\,R^{k-1},
	\qquad 0 < \beta \leq 1
\end{equation}
where $\beta$ is a smoothing factor that trades off responsiveness to load changes against noise sensitivity. In the proposed FCS--MPC scheme, the filtered estimate $R^{k+1}$ obtained from  Eq. (\ref{eq25}), is employed in the load-dependent steady-state duty ratio $d_{\text{ss}}(R^{k+1})$ within the cost function. The steady-state duty reference is derived from the DC equilibrium of the boost converter, accounting for parasitic resistances, inductor non-idealities (resistive and core losses), and a lumped equivalent series resistances (ESR) model for the output capacitor as,
\vspace{-0.1cm}
\begin{equation} \label{eq26}
	d_{\text{ss}}(R) = 1 - \frac{V_{\text{in}}}{V_\mathrm{O}} \left(1 + \frac{R_\mathrm{L} + R_\mathrm{C}}{R}\right)
\end{equation}
where $R_\mathrm{L}$ and $R_\mathrm{C}$ represent the inductor and capacitor ESR respectively. This explicit load-dependent term steers the duty ratio toward the equilibrium value required to sustain \(i_\mathrm{L,ref}^{k+1}\) and \(v_\mathrm{C,ref}^{k+1}\) under the anticipated load $R^{k+1}$, thereby providing inherent feedforward compensation during load transients. To achieve current and voltage regulation with explicit duty-ratio control and a robust load-transient response (TR), the following multi-objective cost function is formulated:
\vspace{-0.2cm}
\begin{equation}  \label{eq27}
	\begin{aligned}
		J(u^{k}, d^{k})
		&= w_{\mathrm{v}}
		\Big(
		v_\mathrm{C,ref}^{k+1}
		- \hat{v}_{\mathrm{C,p}}^{\,k+1}
		\big(u^{k}, d^{k}, R^{k+1}\big)
		\Big)^{2}
		\\
		&\quad
		+ w_{\mathrm{i}}
		\Big(
		i_\mathrm{L,ref}^{k+1}
		- \hat{i}_{\mathrm{L,p}}^{\,k+1}
		\big(u^{k}, d^{k}, R^{k+1}\big)
		\Big)^{2}
		\\
		&\quad
		+ \lambda_{\mathrm{u}}
		\big(u^{k} - u^{k-1}\big)^{2}
		+ \lambda_{\mathrm{d}}
		\Big(
		d^{k} - d_{\mathrm{ss}}\big(R^{k+1}\big)
		\Big)^{2}
	\end{aligned}
\end{equation}
where $w_\mathrm{v}$ and $w_\mathrm{i}$ are voltage and current tracking weights which are initially set to 0. $\lambda_\mathrm{u}$ penalizes switching transitions to reduce commutation losses, and $\lambda_\mathrm{d}$ enforces proximity to the $d_{\text{ss}}(R^{k+1})$. When a sudden load change occurs, the fourth term proactively drives $d^k$ toward the new steady-state operating point, while the voltage and current tracking terms refine the transient trajectory. The optimal control pair is obtained by repeated online enumeration.
\begin{equation}  \label{eq28}
	(u^{k,\mathrm{opt}}, d^{k,\mathrm{opt}}) = \arg\min_{u^k \in \{0,1\},\, d^k \in \mathcal{D}} J(u^{k}, d^{k})
\end{equation}

The resulting $d^{k,\mathrm{opt}}$ is applied to an external carrier-based PWM synchronization modulator operating at switching frequency \(f_{\mathrm{SYN}}\), which generates the gate signal $Q(t_k)$ for the physical boost converter. The binary state $u^{k,\mathrm{opt}}$ is retained in the optimization as it serves as the primary control indicator in \(J\), facilitating accurate CCM/DCM transition. A flowchart of the MPC is shown in Fig.~\ref{fig2}(d), right-plane.  This dual-output architecture ensures both high-fidelity current and voltage regulation with explicit duty-ratio control, enabling superior load-transient rejection compared to conventional practices.

\subsubsection{Computational Environment} An NVIDIA GeForce RTX 4090 GPU with 24 GB GDDR6X VRAM serves as the primary training accelerator, backed by an Intel(R) Core((TM) i9-14900HX (32 CPUs), $\thicksim$2.2 GHz. All training and inference runs are executed within the TensorFlow environment.

\section{Case Study: Hybrid PINN+FCS-MPC in DC-DC Boost Converter} \label{sec4}
To rigorously evaluate the steady-state behavior and TR performance of a DC-DC boost converter supported via the proposed PINN+FCS-MPC scheme, a dual-stage TI-PMLKBOOSTEVM experimental boost module is adopted as the converter under test (CUT). The CUT comprises a monolithic asynchronous TPS55340 (SR\(^\mathrm{1}\)) and a multi-phase synchronous LM5122 (SR\(^\mathrm{2}\)) switching regulator w/ a power range $\approx$ 5-25 W. The SR\(^\mathrm{2}\) features a versatile diode-emulation mode (DEM) and a synchronous operation mode (SOM).
\vspace{-0.4cm}
\begin{table}[ht]
\centering
\caption{Key Parameters \& Operating Conditions of the CUT}
\label{tab:t2}
\setlength{\tabcolsep}{2pt}
 \begin{tabular}{||l|c|l|c||}
\hline
\textbf{Parameter} & \textbf{Value} & \textbf{Parameter} & \textbf{Value} \\
\hline\hline
Input Voltage, $V_{\mathrm{in}}$ & 12 V &
Load Resistance, $R$ & 100 $\Omega$ \\
\hline
Desired Output Voltage, $V_{\mathrm{o}}$ & 24 V &
Inductance, $L$ & 10 $\mu$H \\
\hline
Switching Frequency, $f_{\mathrm{sw}}$ & 600 kHz &
Capacitance, $C$ & 1030 $\mu$F \\
\hline
\end{tabular}
\end{table}

To ensure precise duty-ratio coordination, both regulators are synchronized via an external clock signal \(f_{\mathrm{SYN}}\) instead of fixed internal PWM settings. The \texttt{SYNC} and \texttt{SYNCIN/RT} pins of SR\(^\mathrm{1}\) and SR\(^\mathrm{2}\) are driven by \(f_{\mathrm{SYN}}\) generated by the enhanced PWM (ePWM) module of a TI C2000\texttrademark{} TMS320F28379D microcontroller. The synchronization inputs \(V_{\mathrm{SYN}} \in [0,\,5]~\mathrm{V}\) in SR\(^\mathrm{1}\) and SR\(^\mathrm{2}\) are compatible with the 3.3~V ePWM outputs. The \(f_{\mathrm{SYN}}\) injection satisfies the \(f_{\mathrm{sw}}\in\text{200~kHz–1~MHz}\) operating range and \(\pm 20\%\) tolerance for SR\(^\mathrm{1}\) whereas \(f_{\mathrm{SYN}} \in \{f_{\mathrm{sw}},\,2f_{\mathrm{sw}}\}\) for SR\(^\mathrm{2}\). To avoid frequency foldback and ensure valid measurements, SR\(^\mathrm{1}\) operates above 280~kHz, while SR\(^\mathrm{2}\) is configured in \emph{Master2} mode for simplified gate driving, consistent with the specifications in the device datasheet.

The proposed scheme is executed through a hierarchical offline deployment pipeline as shown in Fig.~\ref{fig2}(e), left-plane. Upon successful trials of the CUT across varied operating conditions, it is utilized for the systematic acquisition of training datasets. Data pre-processing (Step 3) utilizes 3000 samples partitioned 70:15:15 for training, testing, and validation. The TensorFlow-based PINN (Step 4) utilizes an Adam optimizer with a mini-batch size of 64 and early stopping to prevent overfitting. Training converges in $\sim$4 minutes over 5000 epochs. The finalized model is exported (Step 5) in ONNX format for MATLAB R2026a/Simscape (Step 6) integration and experimental verification (Step 7).

PINN's accuracy is shown in Fig.~\ref{fig2}(e), right-plane. The \textsf{error vs sample} plot quantifies the relationship between training density and predictive accuracy. A logarithmic scale is used on the vertical axis. RMSE drops from $4\times10^{1}$ at the start down to $10^{-3}$ over 3000 samples. RMSE slightly fluctuates within $[0.5, 1] \times 10^{3}$; however after the 1000-sample mark, the error decays linearly on this log scale. The narrowing 95\% confidence interval (CI) marked with a light-blue shaded area ensures high-fidelity state estimation for precise FCS-MPC execution. The \textsf{loss vs epochs} plot illustrates the simultaneous minimization of the loss components $\mathcal{L}_{\texttt{data}}$ and $\mathcal{L}_{\texttt{phys}}$. Both show a sharp decline within the first 1000 epochs, eventually plateauing near $10^{-4}$ and confirms data alignment.

Following the data-validation, the learned parameters are embedded directly into the specialized testbed illustrated in Fig.~\ref{fig3}(a). To demonstrate robust duty-ratio control using the TMS320F28379D, Fig.~\ref{fig3}(b) presents the switch node voltages $v_{\mathrm{SW},n}$ ($n \in \{1, 2, 3\}$) at \(f_{\mathrm{SYN}}=250 \ \text{kHz}\) with a $T_{\mathrm{SYN}}= \ $4 $\mu$s for $d \in \{0.3, 0.5, 0.75\}$ (top-bottom), confirming stable external clock synchronization. Simulated results in MATLAB R2026a/Simulink illustrated in Fig.~\ref{fig3}(c), (left-right) confirms CCM$\dashrightarrow$BCM$\dashrightarrow$DCM transitions of the CUT under varying load conditions of 12~$\Omega$ (heavy), 200~$\Omega$ (moderate), and 500~$\Omega$ (light) respectively. The corresponding experimental behavior of the switching regulator SR$^{1}$ under constant-current (CC) load conditions of 0.6~A (heavy), 0.4~A (moderate), and 0.3~A (light) is presented in Fig.~\ref{fig3}(d)-(f) respectively. In each figure, the waveforms of the input voltage $V_{\mathrm{in}}$, output voltage $V_{\mathrm{o}}$, switch-node voltage $V_{\mathrm{sw}}$, and input-ripple+switching noise $V_{\mathrm{ir\text{-}sn}}$ are shown from top to bottom.

With nominal $V_{\mathrm{in}} = 12$~V and $V_{\mathrm{o}} = 24$~V, Fig.~\ref{fig3}(d)-(e) show input and output voltage deviations under heavy and moderate loads, mainly due to increased conduction losses and load transients. In contrast, Fig.~\ref{fig3}(f) exhibits stable voltage regulation at a light load of 0.3~A, with the $V_{\mathrm{sw}}$ waveform confirming DCM operation. The measured input ripple and switching noise at test point TP1 (shown in pink) for SR\(^\mathrm{1}\) reaches 1.07~V peak-to-peak (9\%) at an $f_{\text{sw}} =$ 586.9~kHz.

The load TRs are illustrated in Fig.~\ref{fig3}(g)-(l). Here, baseline: on-chip voltage-mode PWM of SR\(^\mathrm{2}\); PI: factory-programmed compensation network. Fig.~\ref{fig3}(g)-(h) show the output voltage response to a load current step-down from 1~A to 0.2~A, with settling times $\Delta t_{\mathrm{d}}$ of 100~$\mu$s and 48~$\mu$s evaluated around 100~$\mu$s after the transient in both cases. Fig.~\ref{fig3}(i)-(j) present the response to a load current step-up from 0.2~A to 1~A, yielding settling times of 106~$\mu$s and 50~$\mu$s respectively. Next, we employ a DC electronic load (EL) in CR mode, the standard for resistive DC-side characterization. Fig.~\ref{fig3}(k)-(l) depict a resistive load transition from 200~$\Omega$ to 500~$\Omega$, during which $V_{\mathrm{o}}$ increases from 29.8~V to 42~V and the load current follows accordingly. The settling time is reduced from 520~$\mu$s w/o PINN+FCS-MPC assist to 120~$\mu$s w/ the proposed control scheme. In addition, the assisted case exhibits reduced output voltage ripple and faster stabilization. PF applies only at the AC grid-interface, which is beyond the scope of our study.

\vspace{-0.12cm}
\section{Conclusion} \label{sec5}
This study presents a hybrid PINN+FCS-MPC framework for DC-DC boost converters, enabling physics-informed modeling with predictive optimization. By embedding physical laws into the neural network, the controller ensures physically consistent state estimation and robust constraint satisfaction. Experimental results confirm superior performance, including faster transient recovery and reduced voltage ripple across diverse load regulations. This approach provides a computationally efficient, reliable solution for high-performance power electronics, offering a scalable alternative to traditional and purely data-driven control strategies.

\end{document}